\documentclass[aps,prd,amssymb,eqsecnum,showpacs]{revtex4}

\begin{document}

\title{Geometric Boundary Data for the Gravitational Field}
\author{H-O. Kreiss${}^{1,2}$   and J. Winicour${}^{2,3}$
       }
\affiliation{
${}^{1}$ NADA, Royal Institute of Technology, 10044 Stockholm, Sweden\\
${}^{2}$ Max-Planck-Institut f\" ur
         Gravitationsphysik, Albert-Einstein-Institut, 
	  14476 Golm, Germany\\
${}^{3}$ Department of Physics and Astronomy \\
         University of Pittsburgh, Pittsburgh, PA 15260, USA 
	 }

\begin{abstract}

An outstanding issue in the treatment of boundaries in general relativity is
the lack of a local geometric interpretation of the necessary boundary data.
For the Cauchy problem, the initial data is supplied by the 3-metric and
extrinsic curvature of the initial Cauchy hypersurface, subject to
constraints. This Cauchy data determine a solution to Einstein's equations
which is unique up to a diffeomorphism.  Here, we show how three pieces of
unconstrained boundary data, which are associated locally with the geometry
of the boundary, likewise determine a solution of the initial-boundary value
problem which is unique, up to a diffeomorphism. Two pieces of this data
constitute a conformal class of rank-2 metrics, which represent the two
gravitational degrees of freedom. The third piece, constructed from the
extrinsic curvature of the boundary, determines the dynamical evolution of
the boundary.

\end{abstract}


\maketitle

\section{Introduction}

There exists a well posed Cauchy problem for Einstein's equation which has
the important property that local geometric data representing the 3-metric
and extrinsic curvature of the initial Cauchy hypersurface  determine a
spacetime metric $g_{ab}$ which is unique up to diffeomorphism. Presently,
there are two formulations of the initial-boundary value problem (IBVP) which are
strongly well posed, the Friedrich-Nagy formulation~\cite{fn} and the harmonic
formulation~\cite{wpgs,wpe,isol}, but neither provide a local geometric
interpretation of the boundary data. For the harmonic formulation, there exists a
nonlocal (in time) version in which the geometric interpretation of the
boundary data depends upon a background metric constructed from the initial
Cauchy data~\cite{juerg}. As a result,  the boundary data have a geometric 
interpretation which is nonlocal in time. In this work, we show how boundary
data for the gravitational field can be posed  which are locally determined
by the  geometry of the boundary, in the same sense as the Cauchy data. 

In a Cauchy problem, initial data on a spacelike hypersurface ${\cal S}_0$
determine a solution in the domain of dependence ${\cal D}({\cal S}_0)$
(which consists of those points whose past directed characteristics all
intersect ${\cal S}_0$).  In the IBVP, data on a timelike boundary ${\cal
T}$  transverse to ${\cal S}_0$ are used to extend the solution to the domain
of dependence ${\cal D}({\cal S}_0 \cup {\cal T})$. Strong
well-posedness~\cite{klor}  guarantees the existence of a unique solution
which depends continuously on both the initial data and the boundary data. 

The primary application of the gravitational IBVP is the simulation of an
isolated astrophysical system containing neutron stars and black holes. The
standard approach in numerical relativity, as in computational studies of
other hyperbolic systems, is to introduce an artificial outer boundary ${\cal
T}$, which is coincident with the boundary of the computational grid and
whose cross-sections are spheres surrounding the system.  The ability to
compute the details of the gravitational radiation produced by compact
astrophysical sources, such as coalescing black holes, is of major importance
to the success of gravitational wave astronomy. If the simulation of such
systems is not based upon  a strongly well posed IBVP then the results cannot
be trusted in the domain of dependence of the outer boundary. For
comprehensive reviews of the gravitational IBVP see~\cite{brev,sarbtig}.

For hyperbolic systems which are stable under lower order perturbations, the
global solution in the spacetime manifold ${\cal M}$ can be obtained by
patching together local solutions, i.e. the problem can be {\it localized}.
Thus, for purposes of treating the underlying geometrical nature of the
boundary data, it suffices to concentrate on the local problem in the
neighborhood of a point on the boundary. That is the approach taken in this
paper.

In the Friedrich-Nagy formulation, there are three essential pieces of
boundary data which have geometrical or physical significance. One is the
trace $K$ of the extrinsic curvature $K_{ab}$ of the boundary, which
geometrically determines the location of the boundary.   (Note that the
coordinate specification of the location of the boundary is pure gauge
information and does not determine its location in the same geometric sense
that a curve is determined by its acceleration (curvature), given its initial position
and velocity.) Two other pieces of data in the Friedrich-Nagy formulation,
which are related to the gravitational radiation degrees of freedom, are
supplied by a combination of the Weyl tensor components $\Psi_0$ and
$\Psi_4$ in the Newman-Penrose notation~\cite{np}. The remaining boundary
data specify the gauge freedom. 

The Friedrich-Nagy formulation is based upon a symmetric hyperbolic
Einstein-Bianchi system, with evolution variables consisting of an
orthonormal tetrad, the associated connection coefficients and the Weyl
curvature components. Although it differs from the metric based formulations
used in numerical relativity, the requirement of three pieces of geometric
boundary data should be universally applicable.  (Statements found in the
literature that only two pieces of boundary data suffice to specify the
physical or geometrical properties of the gravitational field are misleading.
They are only true when the boundary has been geometrically specified, e.g.
for an $r={\rm const}$ boundary in a background Schwarzschild geometry.)

The outgoing null vector $K^a$ and ingoing null vector $L^a$ used in defining
$\Psi_0$ and $\Psi_4$, respectively, are determined by the unit normal to the
boundary  $N^a$  and a choice of  unit timelike vector $T^a$ tangent to the
boundary according to
\begin{equation}
           K^a=T^a+N^a\, , \quad   L^a=T^a-N^a.
           \label{eq:fnk}
\end{equation}
The choice of direction of $T^a$  represents gauge freedom in this data.
Friedrich and Nagy are careful to point out that this gauge freedom prevents
interpreting $\Psi_0$ and $\Psi_4$ as purely geometric data.

This shortcoming could perhaps be avoided by choosing these null vectors to
be principal null directions of the Weyl tensor. However, in a general
spacetime this would lead to four possible choices which would then have to
be incorporated (in some yet unknown way) into a well posed problem. An
alternative, proposed in~\cite{hjuerg}, is to base the data on the
eigenvectors $V^a$  determined by the trace-free part of the extrinsic
curvature according to
\begin{equation} 
    (  K_{ab} -\frac{1}{3} H_{ab} K ) V^b = \lambda H_{ab} V^b,
 \end{equation} 
where $H_{ab}$ is the intrinsic 3-metric of the boundary. For a spherical
worldtube in Minkowski space, this  picks out a locally preferred timelike
direction $\tilde T^a$. This suggests that the approach might extend  to a
suitably round outer boundary of an isolated system. However, it is again not
clear whether such an approach can be properly incorporated into the
evolution system.

Here we consider geometric boundary data for metric based formulations of the
IBVP. Our main result is that, along with the initial Cauchy data, a
spacetime metric  satisfying Einstein's equations is uniquely determined up
to diffeomorphism by three pieces of boundary data related locally to the
intrinsic metric $H_{ab}$ and extrinsic curvature $K_{ab}$ of the boundary. 
More specifically, two pieces of boundary data  consist of a conformal class
$\{ Q_{ab}\}$ of rank-2 metrics of signature $(0++)$, which represent the two
gravitational degrees of freedom,. The null eigendirection of $\{ Q_{ab}\}$
uniquely determines a flow of streamlines on the boundary. The third piece of
data, which determines the dynamical evolution of the boundary, is a
component of the extrinsic curvature of the boundary picked out by the unit
vector to these streamlines. In Sec.~\ref{sec:data}, we discuss the
underlying geometry and present our main result as a {\it Local Geometric Data
Theorem}.

The demonstration that this local geometric data leads to a strongly well posed  IBVP
is carried out using the harmonic reduction of Einstein's equations to ten wave
equations, as was the method used in establishing the analogous result for
the Cauchy problem~\cite{bruhat}.  In doing so, the three pieces of local
geometric boundary data must be supplemented by seven additional boundary
conditions. Four of these conditions are supplied by the harmonic coordinate
conditions. The other three fix the remaining freedom in the boundary values
of the harmonic coordinates. In Sec.~\ref{sec:pde}, the resulting harmonic IBVP
is reduced to a set of partial differential equations in the frozen
coefficient formalism, which are subject to a combination of Dirichlet and
Neumann boundary conditions.

The demonstration that the strong well-posedness of the frozen coefficient
version of the harmonic IBVP extends to the full quasilinear problem was
given in~\cite{wpgs,wpe} for the case of Sommerfeld boundary conditions. The
application  of these methods to more general partial differential equations
(PDEs) and their application to boundary conditions for isolated systems was
presented in~\cite{isol} . In Sec.~\ref{sec:estimates}, we demonstrate how
this approach can be extended to Dirichlet and Neumann conditions.

The strong well-posedness of the frozen coefficient harmonic  IBVP with local
geometric boundary data is established in Sec.~\ref{sec:wp}. The key idea is
that the full set of boundary conditions can be applied sequentially, similar
to the approach followed in~\cite{wpgs,wpe} except now applied to a set of
Dirichlet and Neumann conditions rather than Sommerfeld conditions.  Our main
result is then established in Sec.~\ref{sec:lgeom}.

Beyond the issue of a local geometrical characterization of the boundary
data, there are other important aspects of the IBVP which remain
unresolved. For the Cauchy problem, it has been shown that a given initial
data set has a maximal development~\cite{gerochbr}.  Two such maximal
developments  corresponding to the same initial data set must be related by a
diffeomorphism. This property, and the related
issue of geometric uniqueness,  have recently been discussed in the context of the IBVP
in~\cite{hjuerg, sarbtig}. For the Cauchy problem, an essential ingredient of geometric
uniqueness is that two solutions of Einstein's equations with the same initial data are
related by a diffeomorphism. Since both solutions
can be transformed to harmonic coordinates without changing the
local geometric data, this result follows from
the uniqueness of the solution in harmonic coordinates. The same argument applies,
at least locally in time, to
the solutions of the IBVP with the same initial-boundary data, as specified in the
{\it Local Geometric Data Theorem}. Another geometric property of the Cauchy
problem is that two diffeomorphic solutions must have diffeomorphic initial data.
It is unlikely that such a strong property holds for the gravitational IBVP. 
Even for a scalar wave equation, the analogous result does not hold since the same
solution can be specified, say, by either Dirichlet or Neumann boundary data.
We do not  address here the question whether it is possible to give a geometric
classification of those initial-boundary data sets which give rise to diffeomorphically
equivalent solutions  of Einstein's equations. However,  our main result that three
pieces of local geometric boundary data, along with the initial data, determine a
geometrically unique solution is important input to the resolution of this question.

When the emphasis is on geometric issues we use abstract tensor indices, e.g.
$v^a$ to denote a vector field, and when the specific spacetime coordinates
$x^\mu= (t,x^i)$ are introduced we use the corresponding coordinate indices,
e.g. $v^\mu= (v^t,v^i)$.

\section{The initial-boundary data}
\label{sec:data}

We begin with a review of the initial data for the Cauchy problem. The
standard treatment of the Cauchy problem introduces a time foliation ${\cal
S}_t$, with future directed unit normal $n_a$. The embedding of ${\cal S}_t$
in the spacetime manifold ${\cal M}$ then gives rise to the decomposition of
the spacetime metric 
\begin{equation}
          g_{ab} = -n_a n_b + h_{ab} ,
\end{equation}
where $h_{ab}$ is the 3-metric intrinsic to  ${\cal S}_t$.  Geometric initial
data are determined by the intrinsic metric $h_{ab}$ and  extrinsic curvature
$k_{ab}=h_a^c \nabla_c n_b$ of the initial Cauchy hypersurface ${\cal S}_0$,
where $\nabla_a$ is the covariant derivative associated with $g_{ab}$. These
data are subject to the Hamiltonian and momentum constraints
\begin{eqnarray}
   2 G^{ab}n_a n_b ={\cal R} - k_{ab} k^{ab} + k^2 &=& 0 ,
    \label{eq:hamc} \\
   h_c^b  G^{a c}n_a = D_b \left( k^{ab} - h^{ab} k \right) &=&0,
     \label{eq:momc}
\end{eqnarray}
where ${\cal R}$ is the curvature scalar and $D_b$ is the covariant
derivative associated with $h_{ab}$.

The remaining initial data necessary to determine a unique spacetime metric
consist of gauge information, i.e. data that affect the solution only by a
diffeomorphism. In the $3+1$ formulation of Einstein's equations, the gauge
freedom in the metric is governed by the choice of an evolution field
\begin{equation}
   t^a =\alpha n^a +\beta^a, \quad \beta^a n_a =0,
\end{equation}
with lapse $\alpha$ and shift $\beta^a$.
The lapse relates the
unit future-directed normal to the time foliation, according to
\begin{equation}
           n_a=-\alpha\nabla_a t.
\label{eq:n}
\end{equation}
The evolution field is transverse but not in general normal to the Cauchy
hypersurfaces so that it determines the shift according to
\begin{equation}
     \beta^a = h^a_b t^b.
\end{equation}

The initial data required for the formulation of a well posed Cauchy problem
depend upon the choice of hyperbolic reduction of Einstein's equations. Here
we consider the hyperbolic reduction associated with harmonic coordinates, as
used in the classic work of Choquet-Bruhat~\cite{bruhat}. Generalized
harmonic coordinates $x^\mu=(t,x^i)=(t,x,y,z)$ are functionally independent
solutions of the curved space scalar wave equation
\begin{equation}
   g^{ab} \nabla_a \nabla_b x^\mu=   -\hat \Gamma^\mu ,
   \label{eq:cswave}
\end{equation}
where $\hat \Gamma^\mu(g,x)$ are harmonic gauge source
functions~\cite{Friedrich}. Thus the harmonic coordinates can be determined
by the initial data
\begin{equation}
        x^\mu|_{{\cal S}_0}=(0, x^i) \, , 
   \quad \partial_t x^\mu|_{{\cal S}_0}=\delta^\mu_t.
        \label{eq:initxdat}
\end{equation}
In terms of the connection coefficients $\Gamma^\mu_{\rho\sigma}$, the
harmonic coordinate conditions are
\begin{equation}
   {\cal C}^\mu :=\Gamma^\mu -\hat \Gamma^\mu =0,
\label{eq:harmcond}
\end{equation}
where
\begin{equation}
     \Gamma^\mu = g^{\rho\sigma}\Gamma^\mu_{\rho\sigma}= 
    -\frac{1}{\sqrt {-g}}\partial_\rho ( \sqrt{-g}g^{\rho\mu} ),
     \quad g=\det g_{\mu\nu}.
\end{equation}
The hyperbolic reduction of the Einstein tensor results from setting
\begin{equation}
     E^{\mu\nu}:= G^{\mu\nu} -\nabla^{(\mu}{\cal C}^{\nu)} 
       +\frac{1}{2}g^{\mu\nu}\nabla_\rho{\cal C}^\rho =0 ,
       \label{eq:creduced}
\end{equation}
where $C^\nu$ is treated as a vector field in constructing the
covariant derivatives. 

When the harmonic conditions (\ref{eq:harmcond}) are satisfied, the principal
part of (\ref{eq:creduced}) reduces to a curved space wave operator acting on
the densitized metric, i.e.
\begin{equation}
     E^{\mu\nu}= \frac{1}{2\sqrt{-g}}g^{\alpha\beta}\partial_\alpha 
   \partial_\beta( \sqrt{-g}g^{\mu\nu})  + \text{lower order terms} =0.
           \label{eq:reduced}
\end{equation}
Thus the harmonic evolution equations (\ref{eq:creduced}) are quasilinear
wave equations for the components of the densitized metric
$\sqrt{-g}g^{\mu\nu}$. The well-posedness of the Cauchy problem for the
harmonic system (\ref{eq:creduced}) follows from known results for systems of
quasilinear wave equations~\cite{bruhat}. Such results are local in time
since there is no general theory for the global existence of solutions to
nonlinear equations.

Constraint preservation results from applying the contracted Bianchi identity
$\nabla_\nu G^{\mu\nu} =0$ to (\ref{eq:creduced}), which leads to the
homogeneous wave equation
\begin{equation}
    \nabla^\nu \nabla_\nu \, {\cal C}^\mu +R^\mu_\nu \,{\cal C}^\nu=0.
  \label{eq:bianchi}
\end{equation}
If the initial data enforce
\begin{equation}
    {\cal C}^\mu |_{{\cal S}_0}= 0 
    \label{eq:c0}
 \end{equation}
 and 
 \begin{equation}   
       \partial_t  {\cal C}^\mu |_{{\cal S}_0}= 0
      \label{eq:ct0}
\end{equation}
then ${\cal C}^\rho=0$ is the unique solution of (\ref{eq:bianchi}). It is
straightforward to satisfy  (\ref{eq:c0})  by algebraically determining the
initial values of $\partial_t g^{\mu t}$ in terms of the  initial values of
$g^{\mu\nu}$ and their spatial derivatives. In order to see how to satisfy
(\ref{eq:ct0}) note that the reduced equations (\ref{eq:creduced}) imply
\begin{equation}
     G^{\mu\nu} n_\nu=n_\nu \nabla^{(\mu}{\cal C}^{\nu)} 
       -\frac{1}{2}n^\mu \nabla_\nu{\cal C}^\nu.
\label{eq:Gn}
\end{equation}
As a result, if
\begin{equation}
     G^{\mu\nu} n_\nu|_{{\cal S}_0} =0,
     \label{eq:hamomc}
\end{equation}
i.e. if the Hamiltonian and momentum constraints are satisfied by the initial
data, and if the reduced equations (\ref{eq:creduced}) are satisfied, then
\begin{equation}
     [n_\nu \nabla^{(\mu}{\cal C}^{\nu)} 
       -\frac{1}{2}n^\mu \nabla_\nu{\cal C}^\nu ]|_{{\cal S}_0} =0.
\label{eq:ndc}
\end{equation}
It is straightforward to check that if ${\cal C}^\mu |_{{\cal S}_0}= 0$ then
(\ref{eq:ndc}) implies (\ref{eq:ct0}).

By standard results, the Hamiltonian and momentum constraints on the initial
data, along with the reduced evolution equations (\ref{eq:creduced}), imply
that the initial conditions (\ref{eq:c0}) and  (\ref{eq:ct0}) required for
preserving the harmonic conditions are satisfied. Conversely, if the
Hamiltonian and momentum constraints are satisfied initially then
(\ref{eq:Gn}) ensures that they will be preserved under harmonic evolution.
In this way, the conditions ${\cal C}^\mu =0$ substitute for the constraints
of the generalized harmonic formulation. This result extends
to the harmonic formulation of the IBVP. If in addition to (\ref{eq:c0}) and
(\ref{eq:ct0}) the harmonic conditions are enforced on the boundary, i.e.
\begin{equation}
    {\cal C}^\mu |_{{\cal T}}= 0 ,
    \label{eq:cb}
 \end{equation}
then ${\cal C}^\mu=0$ is again the unique solution of (\ref{eq:bianchi})
and the Hamiltonian and momentum constraints remain satisfied.

The free initial gauge data in the harmonic formulation consist of the initial
values of the lapse and shift. For simplicity, we set the initial lapse to unity
and the initial shift to zero so that the metric  components satisfy
\begin{equation}
       g^{tt}= -1 , \quad     g^{ti}= 0 , \quad t=0.
       \label{eq:inlsh}
\end{equation}
Along with the initial geometric data $h^{ij}$ and $k^{ij}$, these determine
a unique solution to the Cauchy problem in the harmonic gauge. In geometric
terms,  $h_{ab}$ and $k_{ab}$ determine a solution which is unique up to a
diffeomorphism.

We now formulate the additional geometric data necessary for the IBVP. In the
IBVP, there is another natural decomposition of the metric at the boundary
${\cal T}$,
\begin{equation}
          g_{ab} =N_a N_b + H_{ab},
          \label{eq:bmd}
\end{equation}
where $N_a$ is the unit outward normal and $H_{ab}$ is the 3-metric intrinsic
to  ${\cal T}$.  The boundary ${\cal T}$ intersects the initial Cauchy hypersurface
${\cal S}_0$ at a 2-dimensional edge ${\cal B}_0$. In general, the spacelike
normal $N_a$ to ${\cal T}$ is not orthogonal to the timelike normal $n_a$ to
${\cal S}_0$. As a result, the geometric  initial data must also include the
hyperbolic angle $\Theta_0$ at the edge given by
\begin{equation}
            \sinh \Theta_0  = N_a n^a |_{{\cal B}_0}.
              \label{eq:hangle}
\end{equation}
The initial velocity of the boundary with respect to the initial Cauchy
hypersurface is governed by $\Theta_0$. 

On the 3-dimensional boundary ${\cal T}$, we represent the local geometric data
which encode the two gravitational degrees of freedom by a conformal class
$\{Q_{ab}\}$ of rank 2 metrics of signature $(0++)$ defined by the equivalence
relation $Q_{ab}\equiv \Omega^2 Q_{ab}$, $\Omega>0$.  On the
boundary, $\{ Q_{ab}\}$  picks out an  eigendirection, with null eigenvalue,
i.e. it determines up to extension a non-vanishing vector field $\tilde T^a$
tangent to ${\cal T}$ which satisfies $Q_{ab}\tilde T^b=0$. In turn, 
$\tilde T^a$ determines a flow of streamlines on ${\cal T}$, which is unique
modulo parametrization. We pick the direction of the flow to point
away from ${\cal B}_0$.

In the region of  ${\cal T}$ disjoint from the edge ${\cal B}_0$,  no
additional properties of  $\tilde T^a$ are assumed. In particular, properties
such as the hypersurface orthogonality of $\tilde T^a$ cannot be determined
without reference to a specific $3$-metric on ${\cal T}$. However,  as a
compatibility condition at the edge ${\cal B}_0$, we identify a member $ Q_{ab}$
of  $\{ Q_{ab}\}$ with the intrinsic 2-metric $q_{ab}$ induced on ${\cal B}_0$ by
the initial data $h_{ab}$,
\begin{equation}
      Q_{ab}|_{{\cal B}_0} = q_{ab}|_{{\cal B}_0} .
      \label{eq:comp0}
\end{equation}
As a consequence, $\tilde T^a$ is normal to ${\cal B}_0$.

Although $\tilde T^a$ is not in general hypersurface orthogonal, it is always
possible to introduce local coordinates $(\tau,y^A)$  on the boundary
satisfying
\begin{equation}
        {\cal L}_{\tilde T} \tau=1\, , \quad    {\cal L}_{\tilde T} y^A= 0,
\end{equation}
i.e. by Lie transport along the streamlines of $\tilde T^a$. Independent of
the freedom in the choice of extension for  $\tilde T^a$, in these
coordinates $\tilde T^A=0$ so that $Q_{\tau\tau}=Q_{\tau A}=0$, i.e. the
non-vanishing components are  $Q_{AB}(\tau,y^A)$. Thus the conformal class 
$\{ Q_{ab}\}$ is represented by $Q_{AB} / \sqrt{ \ det(Q_{CD})}$. The gauge
freedom in this coordinate representation is the parametrization $\tau$ of
the streamlines of $\tilde T^a$ (corresponding to the choice of extension)
and the streamline coordinatization $y^A$ (corresponding to the
diffeomorphisms on the factor space of streamlines obtained by identifying
points on each streamline). This local representation allows comparison of
different conformal data sets. Although, such coordinates would be useful in
setting up a $3+1$ evolution problem, they are not useful in the approach
adopted here for the construction of a solution via harmonic coordinates.

The plan here is to construct a metric $g_{ab}$ satisfying Einstein's
equations such that the intrinsic boundary metric $H_{ab}$ obtained from the
$3+1$ boundary decomposition (\ref{eq:bmd})  has the further $2+1$
decomposition 
\begin{equation}
              H_{ab} = -T_a T_b + Q_{ab},
          \label{eq:bmd2}
\end{equation}
where $Q_{ab}$ belongs to $\{Q_{ab} \}$.  A priori to the construction of a
solution, the only condition on $T_a$ is that $T_a \tilde T^a<0$, so that
$H_{ab}$.has signature $(-++)$ and $T_a$ is future directed.
The inverse to the 3-metric
(\ref{eq:bmd2}) can be expressed in the usual form 
\begin{equation}
          H^{ab} = -T^a T^b + Q^{ab} \, , \quad T^a T_a = -1\, ,
    \quad T^a Q_{ab} = 0 \, ,
         \label{eq:bmd2u}
\end{equation}
from which it follows that
\begin{equation}
                  T^a =- \frac {\tilde T^a }{\tilde T^b T_b}.
\end{equation}
Thus, after the construction of a solution, $T^a$ is the future directed unit
timelike vector tangent to the boundary which is geometrically picked out
as an eigenvector of  $\{ Q_{ab}\}$ with null eigenvalue.  

The data $\{ Q_{ab}\}$, along with gauge conditions, are not sufficient to
determine a unique solution.  The remaining geometric data on the boundary
${\cal T}$ are obtained from its extrinsic curvature
\begin{equation}
            K_{ab}=H_a^c \nabla_c N_b.
\end{equation}
In the Friedrich-Nagy formulation of the IBVP, the trace $K =H^{ab}K_{ab}$
forms part of the boundary data. Using the fact that $H_{ab}$ has $(-++)$
signature, Friedrich and Nagy show that when $K$ is expressed in terms of a
boundary defining function it gives rise to a wave equation for that function
which geometrically determines the location of the boundary. For the method
we use here to establish the strong well-posedness of a metric formulation of
the IBVP, there does not appear to be a way to incorporate $K$ into the
boundary data. However, the alternative component
\begin{equation}
        L=(H^{ab}-T^a T^b)K_{ab}
        \label{eq:L}
\end{equation}
does supply the data in the required form. Because $(H^{ab}-T^a T^b)$ also
has $(-++)$ signature, $L$ geometrically determines the location of the
boundary by the same construction used by Friedrich and Nagy.

We now state our main result.

\medskip

\noindent {\bf Local Geometric Data Theorem:}. {\it The Cauchy data $h_{ab}$
and $k_{ab}$ on ${\cal S}_0$, along with edge data $\Theta_0$ on ${\cal
B}_0$  and boundary data $\{Q_{ab}\}$ and $L$ on ${\cal T}$, determine a
metric which satisfies the vacuum Einstein equations (locally in time) such
that ${\cal T}$ has induced metric of the form (\ref{eq:bmd2}) and extrinsic
curvature component (\ref{eq:L}). The solution is unique, up to a
diffeomorphism. All data are assumed to be smooth and compatible.}

\medskip

Here the Cauchy data must satisfy the Hamiltonian and momentum constraints
but the boundary data are constraint free subject to compatibility with the
Cauchy data. The restriction of $\{ Q_{ab}\}$ and $h_{ab}$ to ${\cal B}_0$ are
required to lead to conformally equivalent 2-metrics via (\ref{eq:comp0}), which
fulfills the lowest order compatibility condition. For a $C^\infty$ solution, the
compatibility conditions involve matching all derivatives of the initial data
and boundary data at points on ${\cal B}_0$. This is a complicated
requirement which we assume has been satisfied. Compatibility conditions pose
no restriction on the boundary data in the region of ${\cal T}$ disjoint from
${\cal B}_0$.

Together $\{ Q_{ab}\}$ and $L$ supply three pieces of boundary data which
have the above local geometric interpretation after the construction of a
solution. As for the case of the Cauchy problem, additional data, which
control the gauge degrees of freedom, are necessary to determine a unique
solution. This gauge data depend upon the particular hyperbolic reduction
used to formulate the IBVP. In the formulation of a strongly well posed
harmonic  IBVP,  the Einstein equations reduce to 10 wave equations for the
components of the metric, so that 10 boundary conditions are necessary. In
addition to the 3 pieces of boundary data $\{ Q_{ab}\}$ and $L$, the harmonic
conditions (\ref{eq:harmcond}) supply 4 boundary conditions, as will be
described in Sec.~\ref{sec:pde}. There are 3 more pieces of gauge data on the
boundary which are necessary to specify completely the harmonic coordinate
freedom. These data pin down the values of the harmonic coordinates on the
boundary, as described below. Together with these harmonic coordinate
conditions, the geometric data determine a strongly well posed problem with
a unique solution.

A non-zero value of the hyperbolic angle $\Theta_0$ presents a technical
complication in prescribing the three pieces of harmonic gauge data. However,
the value of $\Theta_0$ can be adjusted to zero by carrying out a Cauchy
evolution in the neighborhood of ${\cal B}_0$ to a new choice of ${\cal
S}_0$, which keeps  ${\cal B}_0$ unchanged. Since the Cauchy problem is
well posed, the initial data for this modified problem depend continuously on
the initial data for the original problem. Consequently, the original IBVP is
strongly well posed if the IBVP for the modified problem with $\Theta_0=0$ is
strongly well posed. In the following, we assume that this has been carried
out. (Otherwise, the technical details in constructing a convenient gauge for
establishing a well posed IBVP become more complicated; cf.~\cite{wpe}  where
the case $\Theta_0\ne 0$ is treated.) Referring to (\ref{eq:hangle}), the
requirement that $\Theta_0=0$ implies
\begin{equation}
        N_a n^a |_{{\cal B}_0} =0 
          \label{eq:zbs}
\end{equation}
so that the compatibility condition (\ref{eq:comp0}) implies
\begin{equation}
          T^a |_{{\cal B}_0}=n^a |_{{\cal B}_0}.
\end{equation}

Since harmonic coordinates are solutions of the curved space scalar wave
equation, they are determined by the initial data (\ref{eq:initxdat}) along
with boundary data for a scalar wave. The boundary data for these coordinates
can be specified in any form which leads to a strongly well posed IBVP. For
our present purpose, we consider homogeneous Dirichlet or Neumann boundary
data. In order to investigate the possible choices, consider Gaussian normal
coordinates $\hat x^\mu = (\hat t, \hat x, \hat x^A)$ tailored to the
boundary at $\hat x=0$ with ${\cal T}$ coordinatized by  $\hat t \ge 0$ and
$\hat x^A =(\hat y, \hat z)$. In these coordinates, the metric has the form
\begin{equation}
      g_{\hat \mu \hat \nu }dx^{\hat \mu} dx^{\hat \nu} = d \hat x^2 
      +H_{\hat I \hat J}dx^{\hat I} dx^{\hat J} , 
     \quad x^{\hat I}  =(\hat t, \hat x^A) 
\end{equation}
in the neighborhood of the boundary. Thus $g^{\hat x \hat t}=  g^{\hat x \hat
A} =0$ on the boundary

Harmonic coordinates $x^\mu =(t,x,x^A)$,   $x^A =(y,z)$, can now be
introduced by solving an IBVP for the scalar wave equation (\ref{eq:cswave}).
On the boundary we prescribe the homogeneous Dirichlet data $x=\hat x=0$, so
that the boundary is given by $x=0$. For the remaining harmonic coordinates we
prescribe the homogeneous Neumann data 
$$    \frac{\partial x^A}{\partial \hat x} =0, 
    \quad  \frac{\partial t}{\partial \hat x}=0, \quad x=0,
$$
so that on the boundary
$$
   g^{xA} = {\partial x \over \partial{\hat x}} 
    \frac{\partial x^A}{\partial \hat x^\alpha} g^{\hat x \hat \alpha} 
   = {\partial x \over \partial{\hat x}} 
   \frac{\partial x^A}{\partial \hat x} g^{\hat x \hat x} =0.
$$
Similarly $g^{xt}=0$ on the boundary, which is consistent with the initial
condition (\ref{eq:zbs}) at the edge ${\cal B}_0$. In summary, we use the
boundary freedom in the choice of harmonic coordinates to set
\begin{equation}
     g^{xt}|_{\cal T} = g^{xA}|_{\cal T} =0 ,\quad  x|_{\cal T}=0.
     \label{eq:bgauge}
\end{equation}

\section{Reduction to PDEs}
\label{sec:pde}

In order to reduce the IBVP to a set of PDEs for the metric with the
initial-boundary data described in Sec.~\ref{sec:data}, we express the
harmonic Einstein equations (\ref{eq:reduced}) in the form
\begin{equation}
    g^{\alpha\beta} \partial_\alpha \partial_\beta (\sqrt{-g} g^{\mu\nu}) 
                = F^{\mu\nu},
 \label{eq:peinst}
\end{equation}
where the forcing $F^{\mu\nu}$ represents lower order terms which do not
enter the principal part. Since the harmonic gauge source functions play no
essential role in establishing well-posedness, we set $\hat \Gamma^\mu=0$.

In the harmonic coordinates constructed in Sec.~\ref{sec:data},  the initial
data at $t=0$, with the gauge conditions (\ref{eq:inlsh}), consist of 
\begin{eqnarray}
        g^{ij} &=&h^{ij}, \quad g^{ti}=0, \quad  g^{tt}=-1, \nonumber \\
        \partial_t  g^{ij} &=&
          -\frac{1}{2} k^{ij}, \quad  \partial_t (\sqrt{-g} g^{ti})
           =-\partial_j(\sqrt{h} h^{ij}),
         \quad \partial_t (\sqrt{-g} g^{tt}) =0 .
         \label{eq:initdat}
\end{eqnarray}
The boundary data at $x=0$ consist of the  gauge data (\ref{eq:bgauge}),
\begin{equation}
        g^{xt} =0, \quad g^{xA}=0 ,
        \label{eq:hgauge}
\end{equation}
and the geometric boundary data consisting of the conformal class $\{Q_{ab}
\}$ and the field $L$, which is the extrinsic curvature component
\begin{equation}
    L=   (H^{ab} - T^a T^b) K_{ab}
                =-\frac{1}{2}\sqrt{g^{xx}} (H^{\mu\nu}
      - T^\mu T^\nu) \partial_x g_{\mu\nu} .
                \label{eq:Ldata}
\end{equation} Here (\ref{eq:hgauge}) supplies three Dirichlet boundary
conditions,  $\{Q_{ab} \}$ supplies two additional Dirichlet conditions and
(\ref{eq:Ldata}) supplies a Neumann condition on a
combination of metric components. Four additional boundary conditions result
from enforcing the harmonic constraints (\ref{eq:cb}) on the boundary, which
take the form
\begin{equation}
        \partial_\mu (\sqrt{-g} g^{\mu\nu})|_{x=0} =0.
        \label{eq:hcond}
\end{equation}

We now formulate the PDEs for the frozen coefficient  version of the problem.
The material in Sec's.~\ref{sec:estimates} and \ref{sec:wp} shows that
the strong well-posedness of this frozen coefficient problem extends to the
quasilinear problem.  Following the approach used in~\cite{wpgs,wpe} for
Sommerfeld boundary conditions, we localize the problem in the neighborhood
of a point $p$ on the boundary and the wave operator in (\ref{eq:peinst}) is 
frozen to its value at $p$,
\begin{equation}
   g^{\alpha\beta}(x_p) \partial_\alpha \partial_\beta.
\end{equation}
By a constant linear transformation of the harmonic coordinates which keeps
the $x$-direction fixed, we can then set $g^{\alpha\beta}(x_p)  =
\eta^{\alpha\beta} $ (the Minkowski metric). In doing so, the $x$-direction
remains aligned with $N^a$ and we can further Lorentz transform the
$t$-direction into the $T^a$ direction picked out by  $\{Q_{ab} \}$, so that
$T^a(x_p)\partial_a =\partial_t$. In these $(t,x,x^A)$
coordinates, with $x^A=(y,z)$, we extend the Minkowski metric to a
neighborhood of $p$ and linearize the equations in terms of the variable
\begin{equation}
         \gamma^{\mu\nu} =\sqrt{-g}g^{\mu\nu} -\eta^{\mu\nu} .
\end{equation}
The system (\ref{eq:peinst}) then takes the frozen coefficient form
\begin{eqnarray}
&(-\partial_t^2+\partial_x^2+\partial_y^2+\partial_z^2)
\pmatrix{\gamma^{tt} & \gamma^{tx} & \gamma^{ty} & \gamma^{tz}\cr
\gamma^{tx} & \gamma^{xx} & \gamma^{xy} & \gamma^{xz}\cr
\gamma^{ty} & \gamma^{xy} & \gamma^{yy} & \gamma^{yz}\cr
\gamma^{tz} & \gamma^{xz} & \gamma^{yz} & \gamma^{zz}\cr} =F,
 \quad x\ge 0,~ t\ge 0 ,
 \label{eq:gameq}
\end{eqnarray}
with forcing matrix $F$.

In the neighborhood of $p$ we require that the data be sufficiently close to
Minkowski data  to allow the iterative construction of a solution to the
quasilinear problem. This can be arranged by considering the rescaled metric
$g'_{\mu\nu} =\lambda^{-2} g_{\mu\nu}$, where $\lambda<<1$ is a positive
constant; cf. p. 262 of~\cite{wald}. Then $L'=\lambda L$ and, in the
stretched coordinates $x'^\mu=x_p^\mu + \lambda^{-1}(x^\mu-x_p^\mu)$, the
transformed metric has components $g'_{\mu'\nu'}(x')
=g_{\mu\nu}(x)=\eta_{\mu\nu}+O(\lambda)$.

In these coordinates, $g_{AB}=Q_{AB}+O(\lambda^2)$ in the neighborhood of $p$
and the conformal boundary data consist of
\begin{equation}
          \tilde Q_{AB}= Q^{-1/2}Q_{AB} , \quad Q=\det Q_{AB} .
\end{equation}
In the linearized approximation, this reduces to
\begin{equation}
       \tilde Q^{AB}-\eta^{AB}= \gamma^{AB} 
     -\frac{1}{2}\eta^{AB} \eta_{CD}   \gamma^{CD},
       \label{eq:cdata}
\end{equation}
(\ref{eq:Ldata}) reduces to
\begin{equation}
      L= -\frac{1}{2}\partial_x (2\gamma^{xx}+ \gamma^{yy}+ \gamma^{zz}) 
         \label{eq:Ldata2}
\end{equation}
and the harmonic constraints reduce to
\begin{equation}
        \partial_\mu \gamma^{\mu\nu} =0.
        \label{eq:gamcons}
\end{equation}

The boundary conditions for the linearized system now take the form
\begin{equation}
        {1\over 2}(\gamma^{yy}-\gamma^{zz})=q_1(t,y,z),
\label{eq:g1}        
\end{equation}
\begin{equation}
       \gamma^{yz} =q_2(t,y,z),
\label{eq:g2}        
\end{equation}
\begin{equation}
     \partial_x \bigg(\gamma^{xx}+{1\over 2}(\gamma^{yy}+\gamma^{zz})\bigg) 
    =q_3(t,y,z),
\label{eq:g3}        
\end{equation}
\begin{equation}
        \gamma^{xt} =0,
\label{eq:g4}        
\end{equation}
\begin{equation}
       \gamma^{xy} =0,
\label{eq:g5}        
\end{equation}
\begin{equation}
      \gamma^{xz} =0 ,
\label{eq:g6}        
\end{equation}
\begin{equation}
    \partial_t\gamma^{tx}+\partial_x\gamma^{xx}+\partial_y\gamma^{xy}
    +\partial_z\gamma^{xz} =0,
\label{eq:g7}        
\end{equation}
\begin{equation}
       \partial_t\gamma^{ty}+\partial_x\gamma^{xy}+\partial_y\gamma^{yy}
    +\partial_z\gamma^{yz} =0,
\label{eq:g8}        
\end{equation}
\begin{equation}
      \partial_t\gamma^{tz}+\partial_x\gamma^{xz}+\partial_y\gamma^{yz}
    +\partial_z\gamma^{zz}=0,
\label{eq:g9}        
\end{equation}
\begin{equation}
       \partial_t\gamma^{tt}+\partial_x\gamma^{tx}+\partial_y\gamma^{ty}
    +\partial_z\gamma^{tz} =0.
\label{eq:g10}        
\end{equation} The Dirichlet data $q_1$ and $q_2$ are determined from the two
conformally  invariant degrees of freedom contained in (\ref{eq:cdata}) The
Neumann data $q_3$ are determined from the data (\ref{eq:Ldata2}) prescribed
by $L$. The Dirichlet conditions (\ref{eq:g4}) -- (\ref{eq:g6}) arise from
the boundary conditions (\ref{eq:hgauge}) on the harmonic coordinates.  The
boundary conditions  (\ref{eq:g7}) -- (\ref{eq:g10}) arise from the harmonic
constraints  (\ref{eq:gamcons}).

\section{  Energy estimates for quasilinear wave problems with Sommerfeld, 
Dirichlet or Neumann boundary conditions}
\label{sec:estimates}

We establish the strong well-posedness of the IBVP for quasilinear wave equations
with Dirichlet and Neumann boundary conditions by an approach similar to that
carried out in~\cite{wpe} for Sommerfeld boundary conditions. We begin by
reviewing how to obtain energy estimates for the Sommerfeld case.

\subsection{ Sommerfeld boundary conditions}
\label{sec:sommbc}

The  energy estimates in Sections 1--4 of~\cite{wpe} established that the solution of the
frozen coefficient version of the harmonic IBVP with Sommerfeld boundary
conditions is unique and depends continuously on the data.  In Appendix A
of~\cite{wpe}, this result was extended to the strong well-posedness of the
quasilinear problem. Here we first consider a slightly simplified version of
the problem treated in~\cite{wpe}. We show that local existence theorems and
energy estimates for second order quasilinear wave problems can be obtained
in the same way as for first order symmetric hyperbolic systems. It all
depends on {\it a priori}  estimates for arbitrarily high derivatives of the
solutions of linear equations with variable coefficients.

Consider the half-plane problem 
\begin{equation}
       u_{tt}=Pu+Ru+F,\quad x\ge 0, \quad t\ge 0, \quad -\infty < y < \infty,
       \label{eq:15}
\end{equation}       
with Sommerfeld-type boundary conditions at $x=0$,
\begin{equation}
\alpha ( u_t+\gamma u)=u_x+q \, ,\quad \alpha >0 \, , \quad \gamma >0
     \hbox { \rm  strictly positive constants} ,
\label{eq:16}
\end{equation}
smooth boundary data $q(t,y)$ and smooth initial data 
\begin{equation}
    u(0,x,y)=f_1(x,y),\quad u_t (0,x,y)=f_2(x,y) .
    \label{eq:17}
\end{equation}
Here the subscripts $(t,x,y,z)$ denote partial derivatives, e.g.
$u_t=\partial_t u$,
$$ Pu=(au_x)_x+(bu_y)_y-2\gamma u_t-\gamma^2 u $$
and
$$
Ru=c_1u_t+c_2u_x+c_3 u_y+c_4 u.
$$
$Ru$ are terms of lower (first and zeroth) differential order. Also, we use
the notation
$$ 
   (u,v), \quad \|u\|^2=(u,u);\quad (u,v)_B,\quad
     \|u\|^2_B=(u,u)_B
$$
to denote the $L_2$ scalar product and norm over the half-plane and 
boundary, respectively.

All coefficients and data are smooth real functions and $a\ge a_0>0,~ b\ge
b_0 >0$, where $a_0,b_0$ are strictly positive constants. The initial data
are compatible with the boundary conditions. In the above, $\gamma >0$ is a constant
obtained by the change of variables $u\to e^{\gamma t}u^\prime$ and then
deleting the ``prime''. This introduces the term $\gamma^2\|u\|^2$ in  the
energy
\begin{equation}
   E := \|u_t\|^2+(u_x,au_x)+(u_y,bu_y)+\gamma^2\|u\|^2,
\label{eq:energy}
\end{equation}
which provides an estimate of $\|u\|^2$.

\medskip

\noindent  {\bf Lemma.} {\it  There is an energy estimate which is stable
against lower order perturbations.}
\par
\medskip\noindent
{\it Proof:} Integration by parts gives
\begin{eqnarray}
\partial_t E &=& \partial_t \left(\|u_t\|^2+(u_x,au_x)
  +(u_y,bu_y)+\gamma^2\|u\|^2\right)  
      \nonumber \\ 
&=& -4\gamma\|u_t\|^2+2(u_t,F)+2(u_t,Ru)-2(u_t,au_x)_B \nonumber \\
& +&a_t\|u_x\|^2+b_t\|u_y\|^2 \nonumber \\
&\le& {\rm const}  (E+\|F\|^2)-2(u_t,au_x)_B .
\label{eq:18}
\end{eqnarray}
Here, and below, the inequalties follow from the basic inequality
$$
   (u,v) \le   \frac{1}{2} (A^2 \|u\|^2 +A^{-2}  \|v\|^2 ).
$$
Using the boundary conditions, we obtain
$$
-(u_t,au_x)_B=  -(u_t, a \alpha u_t)_B -(u_t, a \alpha  \gamma u)_B  
   +(u_t,aq)_B \le -\frac{1}{2} a_0 \alpha \gamma  \partial_t  \|u\|^2_B + 
   {\rm const} (\|u\|^2_B+\|q\|^2_B).
$$
Therefore (\ref{eq:18}) implies 
\begin{equation}
 \partial_t ( E+ a_0 \alpha \gamma \|u\|^2_B )\le {\rm const}
    (E+\|F\|^2+\|u\|^2_B+\|q\|^2_B). 
\label{eq:19}
\end{equation}
This proves the lemma.

\medskip

Now we can estimate the derivatives. Let $v=u_y,~w=u_t$. Differentiation of
the differential equation gives 
\begin{eqnarray}
v_{tt}&=&Pv+Rv+R_y u+(a_yu_x)_x+(b_yv)_y+F_y,\nonumber \\
w_{tt}&= &Pw+Rw+R_tu+(a_tu_x)_x+(b_tv)_y+F_t .
\label{eq:20}
\end{eqnarray}
Here $R_yu$ and $R_tu$ are linear combinations of first derivatives of $u$
which we have already estimated and can be considered part of the forcing.

The differential equation (\ref{eq:15}) tells us that
$$ au_{xx}=w_t-bv_y+ \hbox{ \rm terms we have already estimated.}
$$ Thus $u_{xx}$ is lower order with respect to $v$ and $w$ and, except for
lower order terms, $v$ and $w$ are solutions of the same differential
equation as $u$. They obey the same boundary conditions with data  $q_y(t,y)$
and $q_t(t,y)$, respectively. Therefore we can estimate all second
derivatives. Repeating the process, we can estimate any number of
derivatives.

We can now proceed in the same way as in~\cite{klor}, where we have
considered first order systems to obtain existence  theorems for equations
with variable coefficients. We approximate the differential equation by a
stable difference approximation and prove, using summation by parts, that the
corresponding estimates for the divided differences hold independently of
gridsize. In the limit of vanishing gridsize, we obtain the existence
theorem.   Since we can estimate any number of derivatives, it is well known,
using Sobolev's theorem, that we can obtain similar, although local in time,
estimates for quasilinear systems. By the same iterative methods as for first
order symmetric hyperbolic systems it follows that strong well-posedness
extends locally in time to the quasilinear case, as well as other standard
results such as the principle of {\it finite speed of propagation.}

\medskip
\noindent {\bf Remark.} {\it There are no difficulties to extend the results
to three spatial dimensions.}

\medskip

\subsection{Dirichlet and Neumann conditions}
\label{sec:homdn}

If we replace the Sommerfeld boundary conditions by homogeneous Dirichlet or
Neumann conditions, then the boundary term $(u_t,au_x)_B $ in (\ref{eq:18})
vanishes. Thus the energy estimates in Sec.~\ref{sec:sommbc} clearly
hold for homogeneous Dirichlet or Neumann conditions with boundary data $q=0$.

Now we consider the half-plane problem for wave equations with inhomogeneous
Dirichlet or Neumann boundary conditions.  As we will show, we can
transform these problems into problems with homogeneous boundary conditions 
by changing the forcing and the  initial data. As a model problem, we
consider the half-plane problem
\begin{eqnarray}
u_{tt}&=&\left(a(t,x,y)u_x\right)_x+\left(b(t,x,y)u_y\right)_y +F(t,x,y),
     \label{eq:1.1} \\
& x&\ge 0, \quad t\ge 0, \quad -\infty < y < \infty , \nonumber
\end{eqnarray}
with initial conditions
\begin{equation}
      u(0,x,y)=f_1(x,y),\quad u_t(0,x,y)=f_2(x,y), 
      \label{eq:(1.2}
\end{equation}
and Dirichlet boundary condition
\begin{equation}
     u(t,0,y)=q(t,y). 
     \label{eq:1.3} 
\end{equation}     
We assume that all coefficients and data are compatible and smooth. We make a
change of variable
\begin{equation}
      \tilde u(t,x,y)=u(t,x,y)-\varphi(x)q(t,y).
       \label{eq:1.4}
\end{equation}
Here $\varphi(x)$ is a smooth function, with $\varphi(0)=1,$ which decays
exponentially. Then
\begin{equation}
\tilde u(t,0,y)=0, \quad \hbox {\rm i.e. $\tilde u$ satisfies
homogeneous Dirichlet boundary conditions.}
\label{eq:1.5}
\end{equation}
By (\ref{eq:1.4}),
\begin{eqnarray}
      \tilde u_{tt}&=&u_{tt}-(\varphi(x)q(t,y))_{tt},\nonumber \\
    (a\tilde u_x)_x&=&(au_x)_x-\left(a(\varphi(x)q(t,y))_x\right)_x,
      \label{eq:1.6} \\
(b\tilde u_y)_y&=&(bu_y)_y- \left(b(\varphi(x)q(t,y))_y\right)_y. \nonumber
\end{eqnarray}
Finally, by (\ref{eq:1.1}), (\ref{eq:1.5}) and (\ref{eq:1.6}) we obtain the
differential equation with modified forcing term
\begin{equation}
   \tilde u_{tt}=\left(a(t,x,y)\tilde u_x\right)_x+ 
  \left(b(t,x,y)\tilde u_y\right)_y+ F+\tilde F, 
\label{eq:1.7}
\end{equation}
which satisfies homogeneous Dirichlet boundary conditions. By assumption, $F$
is a smooth function and $\tilde F$ is composed of $a,b,\varphi$ and $q$ and
their first two derivatives. Since derivatives are smooth functions, $\tilde
F$ is also a smooth function. Therefore $\tilde u(t,x,y)$
satisfies the energy estimates arrived at in Sec.~\ref{sec:sommbc}.

Now we consider (\ref{eq:1.1}) with the Neumann boundary condition
\begin{equation}
      u_x(t,0,y)=q(t,y). 
\label{eq:1.8}
\end{equation}
We make again the transformation (\ref{eq:1.4}) but now with
$\varphi_x(0)=1,$ and obtain the corresponding energy estimate.

As an illustration of how the estimates extend to higher derivatives,
consider the half-plane problem (\ref{eq:1.1}) with a homogeneous Dirichlet
boundary condition for $a= b= 1$ (which poses no restriction),
\begin{equation}
 u_{tt}=  u_{xx}+  u_{yy}+F, \quad u(t,0,y)=0.
 \label{eq:1}
\end{equation} 
Since $F(t,x,y)$ and the data $q(t,y)$, $f_1(x,y)$ and $f_2(x,y)$ are smooth functions,
we can obtain energy estimates for the derivatives of $u$ by 
differentiating (\ref{eq:1}). We obtain
\begin{equation}
u_{ytt}=u_{yxx}+u_{yyy}+F_y, \quad u_y(t,0,y)=0,
\label{eq:3}
\end{equation}
\begin{equation}
u_{ttt}=u_{txx}+u_{tyy}+F_t,\\ \quad u_t(t,0,y)=0.
 \label{eq:4}
\end{equation}
As in (\ref{eq:20}), we introduce the variables
\begin{equation}
 v=u_y,\quad w=u_t.
  \label{eq:5} 
\end{equation}
Then (\ref{eq:3}), (\ref{eq:4}) become
\begin{equation}
v_{tt}=v_{xx}+v_{yy}+F_y, \quad v(t,0,y)=0,
 \label{eq:6}
\end{equation}
\begin{equation}
w_{tt}=w_{xx}+w_{yy}+F_t, \quad w(t,0,y)=0.
\label{eq:7}
\end{equation}
Integration by parts then gives us an energy estimate for
\begin{equation}
\|v_t\|^2+\|v_x\|^2+\|v_y\|^2+\|w_t\|^2+\|w_x\|^2+\|w_y\|^2. 
\label{eq:8}
\end{equation}
By (\ref{eq:1}) and (\ref{eq:5}) we obtain
$$
u_{xx}+F=u_{tt}-u_{yy}=w_t-v_y.
$$
Therefore, by (\ref{eq:8}), we obtain a bound for $\|u_{xx}\|^2.$

We obtain a bound for $\|u_{xxx}\|^2$ in the same way by replacing $v$ and $w$ by
\begin{equation}
 v^{(1)}=u_{yy},\quad w^{(1)}=u_{tt}. 
 \label{eq:9}
\end{equation}
Now we obtain the differential equations
\begin{eqnarray}
v_{tt}^{(1)}&=&v_{xx}^{(1)}+ v_{yy}^{(1)}+F_{yy},\quad v^{(1)}(t,0,y)=0,\cr
w_{tt}^{(1)}&=&w_{xx}^{(1)}+ w_{yy}^{(1)}+F_{tt},\quad w^{(1)}(t,0,y)=0,
\label{eq:10}
\end{eqnarray}
and we obtain energy estimates for
\begin{equation}
  \|v_t^{(1)}\|^2+ \|v_x^{(1)}\|^2+ \|v_y^{(1)}\|^2 +\|w_t^{(1)}\|^2
    + \|w_x^{(1)}\|^2+ \|w_y^{(1)}\|^2\ ,
\label{eq:11}
\end{equation}
which we can express in terms of $u$ according to
\begin{equation}
 \|u_{tyy}\|^2+ \|u_{xyy}\|^2+ \|u_{yyy}\|^2 + \|u_{ttt}\|^2+
   \|u_{xtt}\|^2+ \|u_{ytt}\|^2 .
\label{eq:12}
\end{equation} 
By differentiation of (\ref{eq:1}) with respect to $x$,
\begin{equation}
     u_{xxx}=u_{xtt}-u_{xyy}-F_x.
\end{equation}
From (\ref{eq:12}), we already have estimates for $\|u_{xtt}\|^2$ and
$\|u_{xyy}\|^2$. Therefore we also obtain an estimate for $\|u_{xxx}\|^2$.
This process can be continued.

Our result is not restricted to the model problem but is valid in general.
For example, we can replace (\ref{eq:1.1})  by the corresponding half-plane
problem in three spatial dimensions.

\medskip
\noindent {\bf Remark.}
{\it In many problems, surface waves, glancing waves and other waves specific
to the boundary are important.  In that case, there is no energy estimate and
the above technique does not activate these phenomena. Instead, in such
cases, we split the problem into two problems; one with homogeneous boundary
conditions and another where only the boundary conditions do not vanish, i.e.
the forcing and the initial values are zero. The first is covered by the
results in this Section. The second we treat by Fourier-Laplace techniques.
For examples, see~\cite{kpeters,kortpeters}.} 

\section{The strong well-posedness of the IBVP for the harmonic Einstein
equations}
\label{sec:wp}

In Sec.~\ref{sec:lgeom} we establish the strong well-posedness of the
gravitational IBVP for the system (\ref{eq:gameq}) with boundary conditions
(\ref{eq:g1}) -- (\ref{eq:g10}) determined by local geometric boundary data
and harmonic coordinate conditions. In order to illustrate how the estimates in
Sec.~\ref{sec:estimates} apply  we first progress through a sequence of model
problems which illustrate a rich variety of acceptable boundary conditions.

\subsection{Model problem I: The harmonic Einstein equations in one spatial
dimension}

First consider the half-plane problem in the frozen coefficient formalism of
the harmonic Einstein equations for the system of wave equations in one space
variable
\begin{equation}
\bigl( - \partial_t^2+
 \partial_x^2\bigr)
\pmatrix{\gamma^{tt} &\gamma^{tx}\cr \gamma^{tx} & \gamma^{xx}\cr} =F ,
  \quad x\ge 0,~t\ge 0,
\label{eq:2.1}
\end{equation}
with forcing matrix $F$.
In standard notation, we treat the system in the sequential order
\begin{eqnarray}
(1)\quad &{}&\partial_t^2 \gamma^{tx}=
 \partial_x^2\gamma^{tx}+F_1 ,\nonumber \\
(2)\quad &{}& \partial_t^2\gamma^{xx}=
     \partial_x^2 \gamma^{xx}+F_2,
     \label{eq:2.2}\\
(3)\quad & {}&
\partial_t^2\gamma^{tt}=
\partial_x^2\gamma^{tt}+F_3. \nonumber 
\end{eqnarray}
Here $\gamma^{tx}(t,x),~\gamma^{xx}(t,x),~\gamma^{tt}(t,x)$ denote the
dependent variables which we want to determine on the half-plane. The forcing
terms $F_1(t,x),~F_2(t,x),~F_3(t,x)$ are smooth functions of $(t,x)$.

The solution of our problem is determined by the initial data corresponding
to (\ref{eq:initdat}) along with
\begin{eqnarray}
&{}&\hbox{the Dirichlet boundary condition}\quad
 \gamma^{tx}(t,0)=q(t) \nonumber \\
&{}&\hbox{or the Neumann boundary condition}
  \quad \partial_x \gamma^{tx}(t,0)=q(t),
\label{eq:2.3}
\end{eqnarray}
and the harmonic boundary conditions applied in the sequential order 
\begin{eqnarray}
   \partial_t \gamma^{tx}(t,0)+\partial_x\gamma^{xx}(t,0)&=&0,
      \label{eq:2.4x}  \\
   \partial_t \gamma^{tt}(t,0)+\partial_x\gamma^{tx}(t,0)&=&0.
\label{eq:2.4t} 
\end{eqnarray}

We start with the wave equation for $\gamma^{tx}$ with smooth boundary data
(\ref{eq:2.3}) and smooth compatible initial data. By means of the
transformation (\ref{eq:1.4}) in Sec.~\ref{sec:homdn}, we modify the forcing
so that the variables satisfy homogeneous boundary conditions, which we
denote by 
\begin{equation}
       q(t)\equiv 0.
       \label{eq:qhom}
\end{equation}
Then we can estimate $\gamma^{tx}$ and its derivatives on
the boundary, as well as in the interior $x>0$, in terms of the data. 
The problem is strongly well posed and we
can solve the wave equation  for $\gamma^{tx}$.

Next, since $\gamma^{tx}(t,0)$ is a known smooth function, we
use the harmonic boundary condition (\ref{eq:2.4x}) and obtain smooth
Neumann boundary data $\partial_t \gamma^{tx}(t,0)$ for
$\gamma^{xx}$. We again use the
transformation  (\ref{eq:1.4}) so that $\partial_x \gamma^{xx}(t,0)\equiv 0$,
using the notation (\ref{eq:qhom}). The resulting wave problem for
$\gamma^{xx}$ with homogeneous Neumann data is strongly well posed  so that
we can estimate $\gamma^{xx}(t,x)$ and its derivatives.  Finally, we obtain
the same result for $\gamma^{tt}$, using the harmonic boundary condition
(\ref{eq:2.4t}) and the transformation (\ref{eq:1.4}).

\medskip
\noindent {\bf Remark.} {\it Alternatively, instead of  (\ref{eq:2.3}), we could obtain a
strongly well posed problem by prescribing Dirichlet or Neumann data
for $\gamma^{xx}$ and using the harmonic boundary conditions to solve for the remaining
components in the sequential order $(\gamma^{tx},\gamma^{tt})$.}

\subsection{Model problem II: The harmonic Einstein equations in two spatial
dimensions}

Now consider the half-plane problem in frozen coefficient formalism in two
spatial dimensions,
$$
\bigl( -\partial_t^2+
\partial_x^2+
 \partial_y^2 \bigr)
\pmatrix{\gamma^{tt} & \gamma^{tx} & \gamma^{ty}\cr
\gamma^{tx} & \gamma^{xx} & \gamma^{xy}\cr
\gamma^{ty} & \gamma^{xy} & \gamma^{yy}\cr}=F,
 \quad x\ge 0,~t\ge 0,~-\infty<y<\infty,
$$
where $F$ again represents the forcing. The components
$\gamma^{yy},~\gamma^{tx}$ and $\gamma^{xy}$ satisfy Dirichlet or Neumann
boundary conditions. The initial data correspond to (\ref{eq:initdat}).

The harmonic boundary conditions are applied in the sequential order 
\begin{eqnarray}
 \partial_t \gamma^{tx}(t,0,y)+\partial_x\gamma^{xx}(t,0,y)
     +\partial_y \gamma^{xy}(t,0,y)&=&0,
        \label{eq:2.6x} \\
 \partial_t \gamma^{ty}(t,0,y)+\partial_x\gamma^{xy}(t,0,y)
          +\partial_y \gamma^{yy}(t,0,y)&=&0,
           \label{eq:2.6y}  \\
   \partial_t \gamma^{tt}(t,0,y)+\partial_x\gamma^{tx}(t,0,y)
      +\partial_y\gamma^{ty}(t,0,y)&=&0.
       \label{eq:2.6t}  
\end{eqnarray}

We proceed essentially in the same way as for model problem I. We use the
transformation (\ref{eq:1.4}) so that the wave equations for
$\gamma^{yy},~\gamma^{tx}$ and $\gamma^{xy}$ satisfy homogeneous Dirichlet or
Neumann boundary conditions. Then the corresponding wave problems 
are well posed and there are energy estimates for these
variables and their derivatives. 

We use the harmonic boundary conditions to obtain estimates
for the remaining variables. First, the boundary condition (\ref{eq:2.6x}) determines
smooth Neumann boundary data for $\gamma^{xx}$ in terms of previously
estimated quantities. After using the transformation (\ref{eq:1.4}), it
reduces to
\begin{equation}
            \partial_x\gamma^{xx}(t,0,y)\equiv 0
\end{equation}
and the resulting wave problem for $\gamma^{xx}$ is strongly well posed. Thus
we can estimate $\gamma^{xx}(t,x,y)$ and its derivatives. Similarly, the
boundary condition (\ref{eq:2.6y}) determines smooth Dirichlet boundary data for
$\gamma^{ty}(t,0,y)$ in terms of previously estimated quantities. After the
transformation (\ref{eq:1.4}), it reduces to
\begin{equation}
            \partial_t\gamma^{ty}(t,0,y)\equiv 0
\end{equation}
so that the resulting wave problem is strongly well posed and
we can estimate $\gamma^{yy}(t,x,y)$ and its derivatives.
Finally, the boundary condition (\ref{eq:2.6t}) determines Dirichlet boundary data for
$\gamma^{tt}(t,0,y)$ in terms of previously estimated quantities
and we can use the transformation (\ref{eq:1.4}) to
obtain a strongly well posed problem for $\gamma^{tt}$.

\subsection{Model problem III: The harmonic Einstein equations in three
spatial dimensions}
\label{sec:mod3}

We now consider the half-plane problem for the linearized harmonic equations
in three spatial dimensions
\begin{eqnarray}
&\bigl( - \partial_t^2+
\partial_x^2+
 \partial_y^2+\partial_z^2 \bigr )
\pmatrix{\gamma^{tt} & \gamma^{tx} & \gamma^{ty}& \gamma^{tz}\cr
\gamma^{tx} & \gamma^{xx} & \gamma^{xy}& \gamma^{xz}\cr
\gamma^{ty} & \gamma^{xy} & \gamma^{yy}&\gamma^{yz}\cr
\gamma^{tz} & \gamma^{zx} & \gamma^{zy}&\gamma^{zz}\cr }=F,
\label{eq:2.7} \\
&\quad x\ge 0,~t\ge 0,~-\infty<y<\infty,~-\infty <z< \infty, \nonumber
\end{eqnarray}
where $F$ represents the forcing,
$\gamma^{yy},~\gamma^{yz},~\gamma^{zz},~\gamma^{tx},~\gamma^{xy}$ and
$\gamma^{xz}$  satisfy Dirichlet or Neumann boundary conditions and the
initial data correspond to (\ref{eq:initdat}).

The harmonic constraints are applied on the boundary in the sequential order
\begin{eqnarray}
 \partial_t \gamma^{tx} +\partial_x\gamma^{xx}
       +\partial_y \gamma^{xy}+\partial_z\gamma^{xz}&=&0, 
        \label{eq:cx} \\
 \partial_t \gamma^{ty}+\partial_x\gamma^{xy}
         +\partial_y \gamma^{yy}+\partial_z\gamma^{yz}&=&0, 
           \label{eq:cy}  \\
\partial_t \gamma^{tz}+\partial_x\gamma^{xz}
       +\partial_y \gamma^{zy}+\partial_z \gamma^{zz}&=&0,
         \label{eq:cz}  \\
 \partial_t \gamma^{tt}+\partial_x\gamma^{tx}
         +\partial_y\gamma^{ty}+\partial_z\gamma^{tz}&=&0.
           \label{eq:ct}  
\label{eq:2.8}
\end{eqnarray}

We proceed in the same way as in two space dimensions. We use the
transformation (\ref{eq:1.4}) so that the six wave equations for
$\gamma^{yy},~\gamma^{yz},~\gamma^{zz},~\gamma^{tx},~\gamma^{xy}$ and
$\gamma^{xz}$ satisfy  homogeneous Dirichlet or Neumann boundary conditions.
Then there is an energy estimate for these variables and their derivatives. 
We then use the constraints to obtain estimates for the remaining variables. 

The constraints  (\ref{eq:cx}) -  (\ref{eq:cz})  determine Neumann boundary
data for $\partial_x \gamma^{xx}(t,0,y,z)$ and Dirichlet boundary data for
$\gamma^{ty}(t,0,y,z)$ and $\gamma^{tz}(t,0,y,z)$ in terms of the previously
estimated variables.  After using the transformation (\ref{eq:1.4}),  the
resulting wave problems are strongly well posed so that we can estimate
$\gamma^{xx}$,  $\gamma^{ty}$ and $\gamma^{tz}$ and their derivatives. The
constraint (\ref{eq:ct}) then provides Dirichlet data $\gamma^{tt}(t,0,y,z)$
for the remaining variable in terms of previously estimated variables.  After
the transformation (\ref{eq:1.4}), the resulting wave problem for
$\gamma^{tt}$ is strongly well posed. 

\subsection{The harmonic Einstein equations with local geometric data}
\label{sec:lgeom}

Now we turn to the 3-dimensional harmonic Einstein system (\ref{eq:2.7}) with
boundary conditions (\ref{eq:g1}) -- (\ref{eq:g10}) determined by the local
geometric boundary data and harmonic coordinate conditions, as prescribed in
Sec.~\ref{sec:pde}.  After applying the transformation (\ref{eq:1.4}), the
conformal metric data (\ref{eq:g1}) -- (\ref{eq:g2}) reduce to the
homogeneous Dirichlet form
\begin{equation}
       (\gamma^{yy}-\gamma^{zz})(t,0,y,z)\equiv 0 ,
    \quad \gamma^{yz}(t,0,y,z) \equiv 0,
       \label{eq:gdatal}
\end{equation}
and the  extrinsic curvature data $L$ (\ref{eq:g3}) reduce to the homogeneous
Neumann form
\begin{equation}       
          \partial_x (\gamma^{yy}+\gamma^{zz}+2\gamma^{xx}) (t,0,y,z)\equiv 0.
\label{eq:g3h}        
\end{equation}
The boundary gauge data (\ref{eq:g4}) -- (\ref{eq:g6}) are already in the
homogeneous Dirichlet form
\begin{equation}
        \gamma^{tx}(t,0,y,z) =\gamma^{xy}(t,0,y,z)= \gamma^{xz}(t,0,y,z)=0.
\label{eq:g4h}        
\end{equation}
The remaining boundary conditions are supplied by the harmonic constraints 
(\ref{eq:g7}) -- (\ref{eq:g10}).

The situation is similar to model problem III but simpler since the gauge
conditions (\ref{eq:g4h}) are already homogeneous and imply that the
harmonic constraint  (\ref{eq:g7}) has the homogeneous form 
\begin{equation}
      \partial_x\gamma^{xx}(t,0,y,z)=0 ,
\label{eq:gs7h}       
\end{equation}
so that (\ref{eq:g3h}) reduces to 
\begin{equation}       
          \partial_x (\gamma^{yy}+\gamma^{zz}) (t,0,y,z)\equiv 0.
\label{eq:g3hr}        
\end{equation}
Together, the homogeneous boundary conditions (\ref{eq:gdatal}), 
(\ref{eq:g4h}), (\ref{eq:gs7h}) and (\ref{eq:g3hr}) determine  strongly
well posed wave problems for the variables
$(\gamma^{yy}-\gamma^{zz})$,  $\gamma^{yz}$, 
$\gamma^{tx}$, $\gamma^{xy}$, $\gamma^{xz}$, $\gamma^{xx}$ and
$(\gamma^{yy}+\gamma^{zz})$, respectively. Thus we can estimate those
variables and their derivatives. Now we can proceed as in Model problem III
to use the harmonic constraints (\ref{eq:g8}) -- (\ref{eq:g10}) in sequential order to
determine the required estimates for the remaining three independent variables
$\gamma^{ty}$, $\gamma^{tz}$ and $\gamma^{tt}$. This determines
a unique solution to the frozen coefficient problem. Along with the applicability
to the quasilinear problem outlined in Sec.~\ref{sec:estimates}, it
establishes the {\it Local Geometric Data Theorem} proposed in
Sec.~\ref{sec:data}.

\section{Discussion}

We have shown how a conformal class of rank-2
metrics $\{Q_{ab}\}$ and an associated extrinsic curvature component $L$
supply local geometric boundary data for a solution of Einstein's
equations which is unique up to a diffeomorphism. The result was obtained by
introducing harmonic coordinates to formulate boundary conditions
for a strongly well posed IBVP.
This method also broadens the possible
formulation of strongly well posed harmonic IBVPs. The technique
in~\cite{wpgs,wpe}  based upon Sommerfeld conditions has been extended to
include Dirichlet and Neumann conditions, subject to the sequential structure
necessary to enforce the harmonic constraints. For computational
applications, Sommerfeld conditions are most benevolent because they allow
numerical error to leave the grid. It is therefore somewhat discordant with
numerical application that a treatment of the boundary based upon local
geometric data must apparently include at least two Dirichlet conditions,
associated with $\{Q_{ab}\}$, and one Neumann condition associated with
the extrinsic curvature, such as the component $L$. 

There are many options in formulating a suitable combination of Dirichlet,
Neumann and Sommerfeld conditions for a strongly well posed problem, provided
the sequential structure is maintained. However, the only  locally geometric
boundary data allowed by the sequential method used here are $\{Q_{ab}\}$ and $L$.
For example, had we used the trace $K$ of the extrinsic curvature of the boundary
instead of the component $L$ then (\ref{eq:g3hr}) would have been replaced by
\begin{equation}       
     \partial_x  (\gamma^{yy}+\gamma^{zz}-\gamma^{tt} )(t,0,y,z) \equiv 0,
     \label{eq:g3hn}
\end{equation}
which does not fit into the sequential structure for applying the
constraints. It remains an open question whether a different analytic approach
can be used to show that trace $K$ boundary data can replace $L$ in a
strongly well posed harmonic IBVP.

An additional issue of high practical importance is the formulation of a strongly
well posed IBVP for the $3+1$ approach which has historically played a major
role in numerical relativity~\cite{york}. In the $3+1$ formalism, instead of
the 10 wave equations of the harmonic system, Einstein's equations are
reduced to a pair of 6 first order in time equations for $h_{ab}$ and
$k_{ab}$, supplemented by 4 conditions which determine the lapse and shift
Perhaps the geometric insight provided by our results can
shed light on this outstanding problem.
 
%%%%%%%%%%%%%%%%%%%%%%%%%%%%%%%%%%%

\begin{acknowledgments}

%%%%%%%%%%%%%%%%%%%%%%%%%%%%%%%%%%%

We are grateful for numerous discussions with H. Friedrich, which supplied
the catalyst for this work. The research was supported by NSF grants
PHY-0854623 and PHY-1201276 to the University of Pittsburgh.

\end{acknowledgments}

\end{document}